\documentclass[preprint,showpacs,amsmath,amssymb,prd,nofootinbib]{revtex4}
\usepackage{epsf}
\usepackage{dcolumn}
\usepackage{amsmath}
\usepackage{amssymb}
\usepackage{graphicx}

\usepackage[breaklinks=true,colorlinks=true]{hyperref}

\begin{document}

\title{On the hidden symmetries of relativistic hydrogen atom}

\author{Volodimir Simulik\footnote{Email: vsimulik@gmail.com}}

\affiliation{Institute of Electron Physics of Ukrainian National Academy of Scientists,
Uzhgorod 88017, Ukraine}


\begin{abstract}
The Dirac equation in the external Coulomb field is proved to possess the symmetry determined by the 31 operators, which form the 31-dimensional algebra. Two different fermionic realizations of the SO(1,3) algebra of the Lorentz group are found. Two different bosonic realizations of this algebra are found as well. All generators of the above mentioned algebras commute with the operator of the Dirac equation in the external Coulomb field, and, therefore, determine the algebras of invariance of such Dirac equation. Hence, the spin s=(1,0) Bose symmetry of the Dirac equation for the free spinor field, proved recently in our papers, is extended here for the Dirac equation interacting with external Coulomb field. Relativistic hydrogen atom is modeling here by such Dirac equation. We are able to prove for the relativistic hydrogen atom both the fermionic and bosonic symmetries known from our papers about the case of non-interacting spinor field. New symmetry operators were found on the basis of new gamma matrix representations of the Clifford and SO(8) algebras, which were found recently in our papers. Hidden symmetries were found both in the canonical Foldy--Wouthuysen and in the covariant Dirac representations. The symmetry operators, which are simple and graceful in the Foldy--Wouthuysen representation, become non-local in the Dirac model. 
\end{abstract}

\maketitle


\section{Introduction}

Symmetry studies of the equations for the hydrogen atom originate from the non-relativistic case. The SO(4) symmetry of the non-relativistic Schr$\mathrm{\ddot{o}}$dinger equation for a hydrogen atom was found by V. Fock \cite{Fock1}, see also \cite{Barg2}.

Relativistic hydrogen atom is modeling here by the Dirac equation in the external Coulomb field
\begin{equation}
\left(i\partial_{0}-\widehat{H}\right)\psi(x)=0; \quad \widehat{H}\equiv \gamma^{0}\vec{\gamma}\cdot\vec{p}+\gamma^{0}m-\frac{Ze^{2}}{\left|\vec{x}\right|},
\label{Dirac}
\end{equation}
where
\begin{equation}
x\in \mathrm{M}(1,3), \quad \partial_{\mu}\equiv \partial/\partial x^{\mu}, \quad Z=1, \quad \mu=\overline{0,3}, \quad j=1,2,3,
\label{Mink}
\end{equation}
$\mathrm{M}(1,3)=\{x\equiv(x^{\mu})=(x^{0}=t, \, \overrightarrow{x}\equiv(x^{j}))\}$ is the Minkowski space-time and 4-component function $\psi(x)$ belongs to rigged Hilbert space
\begin{equation}
\mathrm{S}^{3,4}\subset\mathrm{H}^{3,4}\subset\mathrm{S}^{3,4*}.
\label{Schwartz}
\end{equation}

Note that due to a special role of the time variable $x^{0}=t\in (x^{\mu})$  (in obvious analogy with non-relativistic theory), in general consideration one can use the quantum-mechanical rigged Hilbert space (\ref{Schwartz}). Here the Schwartz test function space $\mathrm{S}^{3,4}$ is dense in the Schwartz generalized function space $\mathrm{S}^{3,4*}$  and $\mathrm{H}^{3,4}$ is the quantum-mechanical Hilbert space of 4-component functions over $\mathrm{R}^{3}\subset \mathrm{M}(1,3)$.

In order to finish with notations, assumptions and definitions let us note that here the system of units  $\hbar=c=1$ is chosen, the metric tensor in Minkowski space-time $\mathrm{M}(1,3)$ is given by
\begin{equation}
g^{\mu\nu}=g_{\mu\nu}=g^{\mu}_{\nu}, \, \left(g^{\mu}_{\nu}\right)=\mathrm{diag}\left(1,-1,-1,-1\right); \quad x_{\mu}=g_{\mu\nu}x^{\mu},
\label{Metric}
\end{equation}
and summation over the twice repeated indices is implied. The Dirac $\gamma$ matrices are taken in the standard Dirac-Pauli representation
\begin{equation}
\gamma^{0}=\left| {{\begin{array}{*{20}c}
 \mathrm{I} \hfill &  0 \hfill\\
 0 \hfill & -\mathrm{I}  \hfill\\
 \end{array} }} \right|, \quad \gamma^{\ell}=\left| {{\begin{array}{*{20}c}
 0 \hfill &  \sigma^{\ell} \hfill\\
 -\sigma^{\ell} \hfill & 0  \hfill\\
 \end{array} }} \right|, \quad \ell=1,2,3,
\label{Gamm} 
\end{equation}
\noindent where the Pauli matrices are given by
\begin{equation}
\sigma^{1}=\left| {{\begin{array}{*{20}c}
 0 \hfill &  1 \hfill\\
 1 \hfill & 0  \hfill\\
 \end{array} }} \right|, \, \sigma^{2}=\left| {{\begin{array}{*{20}c}
 0 \hfill &  -i \hfill\\
 i \hfill &   0  \hfill\\
 \end{array} }} \right|, \, \sigma^{3}=\left| {{\begin{array}{*{20}c}
 1 \hfill &  0 \hfill\\
 0 \hfill & -1  \hfill\\
\end{array} }} \right|, \, \sigma^{1}\sigma^{2}=i\sigma^{3}, \, 123!-\mathrm{circle}.
\label{Sigm}
\end{equation}

Below we prove the Dirac equation in the external Coulomb field (\ref{Dirac}) to possess the symmetry determined by the 31 operators, which form the 31-dimensional algebra SO(6)$\oplus i\gamma^{0}\cdot$SO(6)$\oplus i\gamma^{0}$. Two different fermionic D(1/2,0)$\oplus$(0,1/2) representations of the SO(1,3) algebra of the Lorentz group are found. Two different bosonic tensor-scalar D(1,0)$\oplus$(0,0) and vector D(1/2,1/2) representations of this algebra are found as well. The corresponding generators of the above mentioned algebras commute with the operator of the Dirac equation in the external Coulomb field (\ref{Dirac}), and, therefore, determine the hidden symmetries (algebras of invariance) of such Dirac equation.

At first we consider the known symmetries of the Dirac equation (\ref{Dirac}), after that we present the mathematical tools, which are necessary for our investigations and, finally, the list of different hidden symmetries of the relativistic hydrogen atom. 

\section{Known symmetries of the Dirac equation in external Coulomb field}\label{Known}

The first four constants of motion (symmetry operators, which commutes with the operator of the Dirac equation) for the equation (\ref{Dirac}) were found by P. Dirac in his paper \cite{Dirac3}, where the equation (\ref{Dirac}) has been derived and introduced. They are three components of the vector $\vec{J}=(J^{1}, \, J^{2}, \, J^{3})$ of the total angular momentum 
\begin{equation}
\vec{J}=\vec{L}+\vec{s}, \quad \vec{L}\equiv \vec{x} \times \vec{p}, \quad \vec{s} \equiv \frac{1}{2}\left| {{\begin{array}{*{20}c}
 \vec{\sigma} \hfill &  0 \hfill\\
 0 \hfill & \vec{\sigma}  \hfill\\
 \end{array} }} \right|,
\label{Yot}
\end{equation}
where $\vec{L}$ is the orbital angular momentum, $\vec{s}$ is the spin-1/2 angular momentum, and the found by Dirac additional constant of motion $K$:
\begin{equation}
K=\gamma^{0}\left(2\vec{s}\cdot\vec{L}+1\right), \quad K^{2}= \vec{J}^{2}+ \frac{1}{4}.
\label{Ka}
\end{equation}

Next symmetry operator is the Johnson--Lippman constant of motion \cite{Jons4}
\begin{equation}
D=2\vec{s} \cdot \frac{\vec{x}}{\left|\vec{x}\right|}+\frac{1}{mZe^{2}}K\gamma^{4}\left(\widehat{H}-\gamma^{0}m\right), \quad D^{2}= 1+\left(\frac{\widehat{H}^{2}}{m^{2}}-1\right)\frac{K^{2}}{Z^{2}e^{4}},
\label{Jonlip}
\end{equation}
which commutes with the operator $\left(i\partial_{0}-\widehat{H}\right)$ of the Dirac equation and anti-commutes with the Dirac symmetry operator $K$ of (\ref{Ka}). Here the anti-Hermitian $\gamma^{4}=\gamma^{0}\gamma^{1}\gamma^{2}\gamma^{3}$ instead of Hermitian $\gamma^{5}$ of other authors is used. Note that article \cite{Jons4} about excellent result (\ref{Jonlip}) was published as a brief 1/10 of a journal page remark, containing only single formula for $D$ from (\ref{Jonlip}). 

After that the way to SO(4) symmetry of the relativistic hydrogen atom was direct. This symmetry was found in \cite{Groot5} and \cite{Stah6}, for the consideration in \cite{Chen7} see the comments of \cite{Stah8}. Thus, the SO(4) symmetry of the Dirac equation (\ref{Dirac}) for the hydrogen atom is given by the six operators 
\begin{equation}
\vec{I}=\vec{J}+\vec{T}, \quad \vec{R}=\vec{J}-\vec{T},
\label{So}
\end{equation}
where $\vec{J}$ is known from (\ref{Yot}) and components of $\vec{T}=(T^{1}, \, T^{2}, \, T^{3})$ have the form
\begin{equation}
T^{1}=\frac{D}{2\sqrt{D^{2}}}, \quad T^{2}=\frac{iDK}{2\sqrt{D^{2}K^{2}}}, \quad T^{3}=\frac{K}{2\sqrt{K^{2}}}. 
\label{Te}
\end{equation}
In the paper \cite{Groot5} the object $\vec{T}=(T^{1}, \, T^{2}, \, T^{3})$ (\ref{Te}) was called as Lentz spin-1/2 vector operator. The notations used in (\ref{Te}) are explained in (\ref{Ka}) and (\ref{Jonlip}) above.

Next symmetry is as follows. The Pauli-G$\ddot{\mathrm{u}}$rsey operators \cite{Pauli9} and \cite{Gurs10}
\begin{equation}
s^{01}=\frac{i}{2}\gamma^{2}\hat{C}, \quad s^{02}=\frac{1}{2}\gamma^{2}\hat{C}, \quad s^{12}=-\frac{i}{2},
\label{Pauli}
\end{equation}
where $\hat{C}$ is the operator of complex conjugation, $\hat{C}\psi=\psi^{*}$ (the operator of involution in the space $\mathrm{H}^{3,4}$), determine  according to \cite{Kri11} the SO(1,2)$\subset$SO(1,3) algebra of invariance of the Dirac equation in the form  $(i\gamma^{\mu}\partial_{\mu}-m+\frac{e^{2}}{\left|\vec{x}\right|})\psi(x)=0$.

In \cite{Nik12} the symmetry of relativistic hydrogen atom in the form of $gl$(8,R) algebra has been found. The stationary case of the Dirac equation (\ref{Dirac}) was considered and the discrete transformation were used.

The author of \cite{Rui13} considered another problem. The quasi-potential two-particle model was presented for the description of spinless relativistic hydrogen atom. The O(4) symmetry and its breaking were investigated.

\section{Briefly on the gamma matrix representations of the real Clifford and SO(8) algebras}\label{Algebras}

In our long time investigations of the mapping of the Maxwell theory on the Dirac theory we also studied the different exotic representations of the Clifford--Dirac algebra, which were useful in the description of the Maxwell theory in the notations taking from the formalism of the Dirac spinor field, see, e.g., \cite{Sim14} and \cite{Sim15}. Step by step we have come to the idea of generalization of the Clifford-Dirac algebra in order to describe the bosonic features of the Dirac theory and fermionic features of the Maxwell theory as well.  

Recently, see, e.g., \cite{Sim16}, we put into consideration the gamma matrix representations of the Clifford algebra $\textit{C}\ell^{\mathbb{R}}$(0,6) and  the Lie algebra SO(8), which are defined over the field of real numbers. Comparison with well known gamma matrix representations of algebras $\textit{C}\ell^{\mathbb{C}}$(1,3) (in physical literature the Clifford--Dirac algebra) and SO(1,5), defined over the field of complex numbers, demonstrates that new representations contain much more useful elements and perspectives for application. Appealing to the gamma matrix representations of the Clifford algebra $\textit{C}\ell^{\mathbb{R}}$(0,6) and  the Lie algebra SO(8) made possible, see, e.g., \cite{Sim17} and \cite{Kri18}, to find the hidden symmetries of the free (non-interacting) Dirac equation. The Bose symmetries were found as well \cite{Sim19}. Below we use these new algebraic objects \cite{Sim16} for the problem of finding the hidden symmetries of the relativistic hydrogen atom. The necessary part of consideration from \cite{Sim16} is presented below in a compact version.  

Consider the fact that seven $\gamma$ matrices
\begin{equation}
\gamma^{1},\,\gamma^{2},\,\gamma^{3},\,\gamma^{4}=\gamma^{0}\gamma^{1}\gamma^{2}\gamma^{3},\,\gamma^{5}=\gamma^{1}\gamma^{3}\hat{C},
\,\gamma^{6}=i\gamma^{1}\gamma^{3}\hat{C},\,\gamma^{7}=i\gamma^{0},
\label{Gen}
\end{equation}
where $\gamma^{\mu}$ matrices are given in (\ref{Gamm}) and operator $\hat{C}$ is defined after the formulas (\ref{Pauli}), satisfy the anti-commutation relations
\begin{equation}
\gamma ^\mathrm{A} \gamma ^\mathrm{B} + \gamma
^\mathrm{B}\gamma ^\mathrm{A} = -
2\delta^{\mathrm{A}\mathrm{B}},\quad \mathrm{A},\mathrm{B}=\overline{1,7},
\label{Anti}
\end{equation}
of the Clifford algebra generators over the field of real numbers. Due to the evident fact that only six operators of (\ref{Gen}) are linearly independent, $\gamma^{4}=-i\gamma^{7}\gamma^{1}\gamma^{2}\gamma^{3}$, it is the representation of the Clifford algebra $\textit{C}\ell^{\mathbb{R}}$(0,6) of the dimension $2^{6}=64$. 

The first 16 operators are given in the Table 1 of \cite{Sim16}, the next 16 are found from them with the help of the multiplication by imaginary unit $i=\sqrt{-1}$. Last 32 are found from first 32 with the help of multiplication by operator $\hat{C}$ of complex conjugation. Thus, if to introduce the notation "stand CD" ("stand" and CD are taken from standard Clifford--Dirac) for the set of 16 matrices from the Table 1 in \cite{Sim16}, then the set of 64 elements of $\textit{C}\ell^{\mathbb{R}}$(0,6) algebra representation will be given by
\begin{equation}
\left\{ ({\mbox{stand \, CD}})\cup i\cdot({\mbox{stand \, CD}})\cup \hat{C}\cdot({\mbox{stand \, CD}})\cup i\hat{C}\cdot({\mbox{stand \, CD}}) \right\}.
\label{Extend}
\end{equation}

As the consequences of the equalities $\gamma^{4}\equiv \prod^{3}_{\mu=0}\gamma^{\mu} \rightarrow \prod^{4}_{\bar{\mu}=0}\gamma^{\bar{\mu}} = -\mathrm{I},$ known from the standard Clifford--Dirac algebra $\textit{C}\ell^{\mathbb{C}}$(1,3), and the consequences of the anti-commutation relations (\ref{Anti}), in $\textit{C}\ell^{\mathbb{R}}$(0,6) algebra for the matrices $\gamma^{\mathrm{A}}$ (\ref{Gen}) the following extended equalities are valid: $\gamma^{7}\equiv -\prod^{6}_{\underline{\mathrm{A}}=1}\gamma^{\underline{\mathrm{A}}} \rightarrow \prod^{7}_{\mathrm{A}=1}\gamma^{\mathrm{A}} = \mathrm{I}, \quad \gamma^{5}\gamma^{6}=i.$

Operators (\ref{Gen}) generate also the 28 matrices 
\begin{equation}
s^{\widetilde{\mathrm{A}}\widetilde{\mathrm{B}}}=\{s^{\mathrm{A}\mathrm{B}}=\frac{1}{4}[\gamma
^\mathrm{A},\gamma
^\mathrm{B}],\,s^{\mathrm{A}8}=-s^{8\mathrm{A}}=\frac{1}{2}\gamma
^\mathrm{A}\},\quad \widetilde{\mathrm{A}},\widetilde{\mathrm{B}}=\overline{1,8},
\label{So}
\end{equation}
which satisfy the commutation relations of the Lie algebra SO(8)
\begin{equation}
[s^{\widetilde{\mathrm{A}}\widetilde{\mathrm{B}}},s^{\widetilde{\mathrm{C}}\widetilde{\mathrm{D}}}]=
\delta^{\widetilde{\mathrm{A}}\widetilde{\mathrm{C}}}s^{\widetilde{\mathrm{B}}\widetilde{\mathrm{D}}}
+\delta^{\widetilde{\mathrm{C}}\widetilde{\mathrm{B}}}s^{\widetilde{\mathrm{D}}\widetilde{\mathrm{A}}}
+\delta^{\widetilde{\mathrm{B}}\widetilde{\mathrm{D}}}s^{\widetilde{\mathrm{A}}\widetilde{\mathrm{C}}}
+\delta^{\widetilde{\mathrm{D}}\widetilde{\mathrm{A}}}s^{\widetilde{\mathrm{C}}\widetilde{\mathrm{B}}}.
\label{Lie}
\end{equation}
It is evident that here we have the algebra over the field of real numbers as well. Furthermore, it is evident that 28 elements (\ref{So}) of SO(8) do not form any Clifford algebra and do not form any subalgebra of the Clifford algebra. It is independent from the Clifford algebra mathematical object. Note that here (as in (\ref{Gen}) for the gamma matrices) the anti-Hermitian realization of the SO(8) operators is chosen, for the reasons see, e.g., the article \cite{Sim20} and monographs \cite{Mon21} or \cite{Mon22}.

The explicit form of the 28 elements of the $\gamma$ matrix representation of the SO(8) algebra is given in the Table 3 of \cite{Sim16}.  The wonderful feature of this gamma matrix representation of the algebra SO(8) is as follows. Here (i) two subsets  ($s^{23},\,s^{31},\,s^{12}$) and ($s^{45},\,s^{64},\,s^{56}$) of operators $s^{\widetilde{\mathrm{A}}\widetilde{\mathrm{B}}}$ from (\ref{So}) determine two different sets of SU(2) spin 1/2 generators, (ii) commute between each other and (iii) commute with the operator of the Dirac equation in the Foldy--Wouthuysen representation \cite{Fold23}.

\section{New symmetries of relativistic hydrogen atom}\label{new}

Below we present both the symmetries of the Foldy--Wouthuysen equation \cite{Fold23} in the external Coulomd field
\begin{equation}
(i\partial _0 - \gamma^{0}\omega +\frac{e^{2}}{\left|\vec{x}\right|})\phi (x)
= 0; \quad \omega \equiv \sqrt{-\Delta + m^{2}}, \; x\in \mathrm{M}(1,3), \; \phi\in \left\{\mathrm{S}^{3,4}\subset\mathrm{H}^{3,4}\subset\mathrm{S}^{3,4*}\right\},
\label{Fol}
\end{equation}
and, of course, the symmetries of the Dirac equation (\ref{Dirac}) in such external field as well. 

The first application of the matrix representations of the algebras $\textit{C}\ell^{\mathbb{R}}$(0,6) and SO(8) is the symmetry analysis (the search of groups and algebras with respect to which the equation is invariant). It is easy to understand that the Foldy--Wouthuysen representation \cite{Fold23} is preferable for such analysis. Indeed, in this representation one must calculate the commutation relations of possible pure matrix symmetry operators from (\ref{So}) only with two elements of the Foldy--Wouthuysen equation (\ref{Fol}) operator: $\gamma^{0}$ and $i$. After the determining of the symmetries of the Foldy--Wouthuysen equation one can find the symmetries of the Dirac equation on the basis of the inverse Foldy--Wouthuysen transformation \cite{Fold23}. Note that after such transformation only the small part of symmetry operators will be pure matrix, the main part of operators will contain the nonlocal operator $\omega \equiv \sqrt{-\Delta + m^{2}}$ and the functions of it.

Now we can start the consideration of the new symmetries of relativistic hydrogen atom. The beginning was suggested in \cite{Sim24}.The fundamental assertions are as follows.

(i) The gamma matrix representation of the subalgebra SO(6) of the algebra SO(8), which is formed by the operators
\begin{equation}
\{s^{\breve{\mathrm{A}}\breve{\mathrm{B}}}\}=\{s^{\breve{\mathrm{A}}\breve{\mathrm{B}}}\equiv\frac{1}{4}[\gamma^{\breve{\mathrm{A}}},\gamma^{\breve{\mathrm{B}}}]\}, \quad \breve{\mathrm{A}},\breve{\mathrm{B}}=\overline{1,6},
\label{Solit}
\end{equation}
determines the algebra of invariance of the Dirac equation in the Foldy--Wouthuysen representation $(\partial _0 +i\gamma^{0}\omega -\frac{e^{2}}{\left|\vec{x}\right|})\phi (x)
= 0$ (in (\ref{Solit}) the six matrices $\{\gamma^{\breve{\mathrm{A}}}\}=\{\gamma^{1},\gamma^{2},\gamma^{3},\gamma^{4},\gamma^{5},\gamma^{6},\}$ are known from (\ref{So})). 

(ii) On the basis of SO(6) (\ref{Solit}) the 31-dimensional gamma matrix representation of the Lie algebra SO(6)$\oplus i\gamma^{0}$SO(6)$\oplus i\gamma^{0}$ is constructed, which is formed by the elements from $\textit{C}\ell^{\mathbb{R}}$(0,6) and is the maximal pure matrix algebra of invariance of the Dirac equation in the Foldy--Wouthuysen representation $(\partial _0 +i\gamma^{0}\omega -\frac{e^{2}}{\left|\vec{x}\right|})\phi (x) = 0$.

(iii) The Dirac equation in external Coulomb field (\ref{Dirac}) is invariant with respect to the 31-dimensional gamma matrix representation of the algebra $\widetilde{\mathrm{SO}}(6)\oplus i\widetilde{\gamma}^{0}\widetilde{\mathrm{SO}}(6)\oplus i\widetilde{\gamma}^{0}$, where the representation of the algebra $\widetilde{\mathrm{SO}}(6)$ is given in the form of (\ref{Solit}) with gamma operators found from matrices (\ref{Gen}) by the inverse Foldy--Wouthuysen transformation \cite{Fold23}. Resulting operators are given by
\begin{equation}
\overrightarrow{\widetilde{\gamma}}=\overrightarrow{\gamma}\frac{-\overrightarrow{\gamma}\cdot\nabla+m}{\omega}+\overrightarrow{p}\frac{-\overrightarrow{\gamma}\cdot\nabla+\omega+m}{\omega(\omega+m)}, \quad \widetilde{\gamma}^{4}=\gamma^{4}\frac{-\overrightarrow{\gamma}\cdot\nabla+m}{\omega},
\label{Image}
\end{equation}
$$\widetilde{\gamma}^{5}=\widetilde{\gamma}^{1}\widetilde{\gamma}^{3}\widetilde{C},
\quad \widetilde{\gamma}^{6}=i\widetilde{\gamma}^{1}\widetilde{\gamma}^{3}\widetilde{C},
\quad \widetilde{\gamma}^{7}=i\widetilde{\gamma}^{0}, \quad \widetilde{\gamma}^{0}=\gamma^{0}\frac{-\overrightarrow{\gamma}\cdot\nabla+m}{\omega},$$
where $\widetilde{C}=(\mathrm{I}+2\frac{i\gamma^{1}\partial_{1}+i\gamma^{2}\partial_{2}}{\sqrt{2\omega(\omega+m)}})\hat{C},$ and $\omega\equiv\sqrt{-\triangle+m^{2}}$. These formulas give the images of gamma matrices (\ref{Gen}) in the Dirac representation after fulfilling the inverse Foldy--Wouthuysen transformation.

Thus, the $\widetilde{\mathrm{SO}}(6)\oplus i\widetilde{\gamma}^{0}\widetilde{\mathrm{SO}}(6)\oplus i\widetilde{\gamma}^{0}$ algebra is found from SO(6)$\oplus i\gamma^{0}$SO(6)$\oplus i\gamma^{0}$ on the basis of the inverse Foldy--Wouthuysen transformation \cite{Fold23}. For the Dirac equation only the part of this algebra is pure matrix, other elements contain the operator $\omega \equiv \sqrt{-\Delta + m^{2}}$.

Consider the symmetries of the relativistic hydrogen atom with respect to the Lorentz group. On the basis of $\textit{C}\ell^{\mathbb{R}}$(0,6) and SO(8) we can determine two different realizations of the $\mathrm{D}(0,\frac{1}{2})\oplus(\frac{1}{2},0)$ representation of the Lie algebra of \textit{universal covering} $\mathcal{L}$ = SL(2,C) of the proper ortochronous Lorentz group $\mbox{L}_ + ^\uparrow $ = SO(1,3)=$\left\{\Lambda=\left(\Lambda^{\mu}_{\nu}\right)\right\}$, with respect to which the equation $(\partial _0 +i\gamma^{0}\omega -\frac{e^{2}}{\left|\vec{x}\right|})\phi (x) = 0$ is invariant:
\begin{equation}
{s}_{\mathrm{I}}^{\mu \nu } = \{{s}_{\mathrm{I}}^{0k} =
\frac{i}{2}\gamma^{k}\gamma^{4},\quad\mbox{ }{s}_{\mathrm{I}}^{km} =
\frac{1}{4}[\gamma^{k},\gamma^{m}]\},\quad\mbox{ }\gamma^{4}
\equiv \gamma^{0}\gamma^{1}\gamma^{2}\gamma^{3},\quad\mbox{ }(k,m
= \overline {1,3}),
\label{Lor}
\end{equation}
\begin{equation}
{s}_{\mathrm{II}}^{\mu \nu } = \{{s}_{\mathrm{II}}^{01} =
-\frac{i}{2}\gamma^{2}\hat{C}, \, {s}_{\mathrm{II}}^{02} = 
-\frac{1}{2}\gamma^{2}\hat{C}, \, {s}_{\mathrm{II}}^{03} = 
\frac{1}{2}\gamma^{0}, \, {s}_{\mathrm{II}}^{23} =
-\frac{1}{2}\gamma^{0}\gamma^{2}\hat{C}, \, {s}_{\mathrm{II}}^{31} =
\frac{i}{2}\gamma^{0}\gamma^{2}\hat{C}, \, {s}_{\mathrm{II}}^{12} =
-\frac{i}{2}\},
\label{Lore}
\end{equation}
Taking the combinations of operators (\ref{Lor}), (\ref{Lore}) we construct the generators of bosonic representations:
\begin{equation}
{s}_{\mathrm{TS}}^{\mu \nu} = \{{s}_{\mathrm{TS}}^{0k} = {s}_{\mathrm{I}}^{0k} +
{s}_{\mathrm{II}}^{0k},\quad {s}_{\mathrm{TS}}^{mn} = {s}_{\mathrm{I}}^{mn} +
{s}_{\mathrm{II}}^{mn}\},\quad\mbox{ } {s}_{\mathrm{V}}^{\mu \nu} = \{
{s}_{\mathrm{V}}^{0k} = -{s}_{\mathrm{I}}^{0k} +
{s}_{\mathrm{II}}^{0k},\quad {s}_{\mathrm{V}}^{mn}={s}_{\mathrm{TS}}^{mn} \},
\label{Bos}
\end{equation}
where ${s}_{\mathrm{TS}}^{\mu \nu}$ and ${s}_{\mathrm{V}}^{\mu \nu}$ are the generators of the tensor-scalar $\mathrm{D}(1,0)\oplus(0,0)$ and irreducible vector $\mathrm{D}(\frac{1}{2},\frac{1}{2})$ representations of the Lie algebra SO(1,3) of the Lorentz group $\mathcal{L}$ respectively, with respect to which the Foldy--Wouthuysen equation $(\partial _0 +i\gamma^{0}\omega -\frac{e^{2}}{\left|\vec{x}\right|})\phi (x) = 0$ is invariant.

Anti-Hermitian operators of every set (\ref{Lor}), (\ref{Lore}), or (\ref{Bos}), satisfy the commutation relations of the Lie algebra SO(1,3) of the Lorentz group $\mathcal{L}$:
\begin{equation}
\left[s^{\mu\nu},s^{\rho\sigma}\right]=-g^{\mu\rho}s^{\nu\sigma}-g^{\rho\nu}s^{\sigma\mu}-g^{\nu\sigma}s^{\mu\rho}-g^{\sigma\mu}s^{\rho\nu}.
\label{Com}
\end{equation}

For the Dirac equation in the space of Dirac spinors $\left\{\psi\right\}$ (i.e. in the Pauli--Dirac representation) the form of the generators of the tensor-scalar D(1,0)$\oplus$(0,0) and irreducible vector D$(\frac{1}{2},\frac{1}{2})$ representations of the Lie algebra SO(1,3) of the Lorentz group $\mathcal{L}$ is similar to (\ref{Bos}) (with (\ref{Lor}), (\ref{Lore})) but the gamma operators are in this case too much complicated and are given by (\ref{Image}). The images of the operators (\ref{Lor}), (\ref{Lore}), or (\ref{Bos}), in the Dirac representation satisfy the commutation relations (\ref{Com}) as well.

In \cite{Sim15} and \cite{Sim19} for the free Dirac and Foldy--Wouthuysen equations we used also the \textit{evident bosonic representation} of (\ref{Bos}), in which the Casimir operators are diagonal and the proof of Bose properties is most convenient. Such \textit{evident bosonic representation} of (\ref{Bos}) can be useful here as well. The corresponding transition operator is given by
\begin{equation}
W = \frac{1}{\sqrt{2}}\left|
\begin{array}{cccc}
 \sqrt{2} & 0 & 0 & 0\\
 0 & 0 & i\sqrt{2}\hat{C} & 0\\
0 & -\hat{C} & 0 & 1\\
0 & -\hat{C} & 0 & -1\\
\end{array} \right|, \quad W^{-1}=\frac{1}{\sqrt{2}}\left|
\begin{array}{cccc}
 \sqrt{2} & 0 & 0 & 0\\
 0 & 0 & -\hat{C} & -\hat{C}\\
0 & i\sqrt{2}\hat{C} & 0 & 0\\
0 & 0 & 1 & -1\\
\end{array} \right|,
\label{Tran}
\end{equation}
$$WW^{-1}=W^{-1}W=\mathrm{I}_{4},$$
and translates the ${s}_{\mathrm{TS}}^{mn}$ operators from (\ref{Bos}) into the form
\begin{equation}
\breve{s}^{1}= \frac{1}{\sqrt{2}}\left|
\begin{array}{cccc}
 0 & 0 & i\hat{C} & 0\\
 0 & 0 & -\hat{C} & 0\\
-i\hat{C} & \hat{C} & 0 & 0\\
0 & 0 & 0 & 0\\
\end{array} \right|, \quad \breve{s}^{2}= \frac{1}{\sqrt{2}}\left|
\begin{array}{cccc}
 0 & 0 & \hat{C} & 0\\
 0 & 0 & -i\hat{C} & 0\\
-\hat{C} & i\hat{C} & 0 & 0\\
0 & 0 & 0 & 0\\
\end{array} \right|, \quad \breve{s}^{3}= \left|
\begin{array}{cccc}
 -i & 0 & 0 & 0\\
 0 & i & 0 & 0\\
0 & 0 & 0 & 0\\
0 & 0 & 0 & 0\\
\label{Ten}
\end{array} \right|,
\end{equation}
$$\overrightarrow{\breve{s}}^{2} =
-1(1+1)\left| {{\begin{array}{*{20}c}
 \mathrm{I}_{3} \hfill & 0 \hfill \\
 0 \hfill & {0} \hfill \\
\end{array} }} \right|,$$
where the Bose character of the operators is evident (here operator of complex conjugation $\hat{C}$ is not matrix).

\section{Discussion and summary}\label{discussion}

The generalization of our results on the Bose symmetries of the free non-interacting Dirac equation (see, e.g., \cite{Sim15} and the references therein)  for the case of the presence of external Coulomb field is suggested. The hidden Fermi and Bose symmetries of the relativistic hydrogen atom are found.

Note that physical picture of the hydrogen atom as the electron in the external Coulomb field is related to the spin 1/2 Fermi symmetries. On the other hand the physical picture of the hydrogen atom as a compound system of proton and electron is related to the total spin 1 (or zero) Bose symmetries.

The main result of this paper is putting into consideration of two different bosonic tensor-scalar D(1,0)$\oplus$(0,0) and vector D(1/2,1/2) representations of the Lie algebra SO(1,3) of the Lorentz group $\mathcal{L}$, with respect to which the Dirac equation in the external Coulomb field is invariant. Two fermionic symmetries of this equation, given by different D(1/2,0)$\oplus$(0,1/2) representations of the SO(1,3) algebra of the Lorentz group, are found as well. The maximal pure matrix symmetry of the Foldy--Wouthuysen equation for hydrogen atom is found and the explicit forms of operators of corresponding symmetry of the Dirac model are calculated.       

\section*{Acknowledgments}

The author is grateful to Igor Gordievich for the participation in few cumbersome calculations.





\end{document}